\documentclass[10pt,aps,prd]{revtex4-2}
\usepackage[a4paper]{geometry}
\usepackage{graphicx}
\usepackage{subfig}
\usepackage[dvipsnames]{xcolor}
\definecolor{custom-blue}{RGB}{6,69,173}
\usepackage{amsfonts}
\usepackage{amsmath}

\usepackage{hyperref}
\hypersetup{colorlinks=true,linkcolor=blue,urlcolor=custom-blue,{citecolor=blue}}
\newcommand{\comm}[1]{}
\def\first_coord{y}
\def\second_coord{z}
\def\proper_time_coord{t}
\def\dimensionless_time_coord{\tau}
\def\rescaled_master_function{\Phi}
\def\stress_energy_tensor{T}
\def\elle{{\dimensionless_time_coord_0}}
\allowdisplaybreaks
\begin{document}
\title{Cosmologies with a magnetic field,  dust, and \texorpdfstring{$\Lambda$}{Lambda}}
\author{M. \v{Z}ofka}
\author{K. Rosquist}
\affiliation{Institute of Theoretical Physics, Charles University}
\affiliation{Department of Physics, Stockholm University}
\email{martin.zofka@matfyz.cuni.cz}
\email{kr@fysik.su.se}
\begin{abstract}
We investigate exact solutions of the Einstein-Maxwell equations with the cosmological constant where the source of the gravitational field consists of a magnetic field and dust. In particular, we restrict our study to the case of Bianchi type III models. All these solutions either start with a singularity and then expand, or they are initially collapsing and end at a singularity. We discuss the physical meaning of the parameters appearing in the metrics and examine the possible subcases and the relative importance of the dust and the magnetic field throughout the evolution of the solution. 
\end{abstract}
\maketitle
\section{Introduction}
It has been known for a long time that there are large-scale magnetic fields present in the universe. The current observations provide both upper \cite{Durrer+Neronov,Paoletti+Chluba+Finelli+Rubino-Martin_2022} and lower \cite{Durrer+Neronov,Vovk+Korochkin+Neronov+Semikoz,lower_bound_on_intergalactic_magnetic_fields} bounds on their strength with the most recent Planck data \cite{Planck2015_constraints_on_primordisal_magnetic_fields} suggesting an upper bound of $10^{-9}$ G for intergalactic magnetic fields of possibly cosmological origin (see \cite{Grasso+Rubinstein} for a review discussing the interactions between magnetic fields and matter in the early universe). It is even possible to constrain the early universe scenarios using the current observations of large-scale magnetic fields \cite{Papanikolaou+Gourgouliatos}. Therefore, it is of interest to consider the evolution of primordial magnetic fields in mutual interaction with both matter and spacetime itself---to solve self-consistently both Einstein and Maxwell equations.

We focus here on generally time-dependent, spatially homogeneous spacetimes with a positive cosmological constant to be consistent with the present interpretations of the cosmological observations. Restricting to Kantowski-Sachs and Bianchi models, there are many solutions in this class where the gravitational field is due to a pressureless dust \cite{Kantowski+Sachs,Ellis,Gron+Eriksen,Gair} or perfect fluid \cite{Kantowski,Kompaneyets+Chernov,Collins,Weber}. It can also be generated by an electromagnetic field alone \cite{Stewart+Ellis}. Some of these solutions also involve the cosmological constant. Some of them are static and some evolve in time, representing thus a cosmological model. For an overview of these works, we refer the reader to \cite{Griffiths+Podolsky,Stephani_Kramer_MacCallum_Hoenselaers_Herlt_2003}. A combination of two or more independent fluids interacting through their gravitational field has also been studied \cite{Coley_Tupper_1986,Faraoni_2021}. Other papers presented solutions involving both a magnetic field and dust or fluid \cite{Thorne,Vajk+Eltgroth,Doroshkevich,Shikin,Alekseev,Tomimura+Waga}.

Other related papers work with, for instance, charged perfect fluid but the 4-current is either spacelike or vanishing \cite{Singh1980}; or the solution is only given implicitly in terms of a differential equation \cite{Casadio_2023}; or the authors simply generate a solution by choosing one of the metric functions and solving for the others, without checking the energy conditions \cite{Ram}; or they fix the spacetime evolution and only solve Maxwell equations on this background \cite{Pal+Maity+Do}.

In the present paper we do include the cosmological term and investigate gravitational fields generated by a pressureless dust in conjunction with a magnetic field, both evolving according to Einstein-Maxwell equations. We study a time-dependent solution assuming it is Bianchi type III. There is an impressive series of works from the 1980s by D. Lorenz who dealt with gravitational fields due to fluids and magnetic fields but---with our assumptions in the present paper---only one of his results \cite{Lorenz} is relevant here. Generally, Lorenz does not dwell so much on the physics of his solutions and rather just lists the analytical results in a given class. We present here a more general exact solution and also analyze in detail the meaning of all the parameters appearing in the metric.

The paper is organized as follows: in Section (\ref{The Ansatz and Einstein equations}), we introduce our ansatz on the metric and other assumptions and solve Maxwell and Einstein equations up to an integral. In Section (\ref{The physics of the solution}), we discuss a number of subcases by switching the individual parameters of the solution on and off and studying the resulting spacetimes in corresponding subsections. Section (\ref{The general case}) explores the properties of the most general case based on the integral appearing in the solution. It also interprets the integration constants that arise when solving the field equations. We conclude in Section (\ref{Conclusions}) with some final remarks and thoughts about possible generalizations and future work. Appendix (\ref{Appendix}) gives the explicit form of the most general case of the central integral appearing in the solution.
\section{The ansatz and Einstein equations}\label{The Ansatz and Einstein equations}
We begin from the general metric of Bianchi type III with $\Lambda$, following \cite{MacCallum_et_al} (which only dealt with vacuum solutions) and also the general metric of (14.13) in \cite{Stephani_Kramer_MacCallum_Hoenselaers_Herlt_2003} but see also section 12.3.1 in \cite{Griffiths+Podolsky}:
\begin{equation}
\mathrm{d}s^2 = - \mathrm{d}\proper_time_coord^2 + X(\proper_time_coord)^2 \mathrm{d}x^2 + Y(\proper_time_coord)^2 \left( \mathrm{d}\first_coord^2 + \sinh^2 \! \first_coord \; \mathrm{d}\second_coord^2 \right),
\end{equation}
with $X(\proper_time_coord)$ and $Y(\proper_time_coord)$ arbitrary functions. The time coordinate $\proper_time_coord$ (the proper time of comoving observers), the spatial coordinate $x$, and the function $Y(\proper_time_coord)$ have the dimension of length while the coordinates $\first_coord$ and $\second_coord$ and the function $X(\proper_time_coord)$ are dimensionless. The metric is of Petrov type D. We seek a general solution involving a pressureless fluid, i.e. dust, which is static in the above coordinate system, and an electromagnetic field. To simplify our equations so as to be able to find an exact solution, we assume a magnetic field aligned with the axis spanned by the coordinate $x$ with only the $F_{yz} = - F_{zy}$ components non-vanishing in the field tensor. The sourceless Maxwell equations
\begin{eqnarray}
\frac {\partial }{\partial t} F_{yz} \left( t,x,y,z \right) = \frac {\partial }{\partial x} F_{yz} \left( t,x,y,z \right) = \frac {\partial }{\partial z} F_{yz} \left( t,x,y,z \right) &=& 0,\\
F_{yz} \left( t,x,y,z \right) \cosh y - \sinh y \frac {\partial }{\partial y} F_{yz} \left( t,x,y,z \right) &=& 0
 \end{eqnarray}
then force a simple form of the electromagnetic field tensor
\begin{equation}\label{electromagnetic_field}
  F = M \sinh \first_coord \:\mathrm{d}\first_coord \! \wedge \! \mathrm{d}\second_coord,
\end{equation}
where the constant of integration $M$ has the dimension of length and determines the strength of the magnetic field observed by comoving observers as $B = M/Y(\proper_time_coord)^2$. The Einstein equations are
\begin{equation}\label{Einstein_xx}
  G_{\mu\nu} + \Lambda g_{\mu\nu} = 8\pi \stress_energy_tensor_{\mathrm{\scriptstyle(elmag)}\mu\nu} + 8\pi \stress_energy_tensor_{\mathrm{\scriptstyle(dust)}\mu\nu},
\end{equation}
with all off-diagonal terms automatically satisfied and $\Lambda$ the cosmological constant. The remaining field equations are
\begin{align}
-\Lambda X Y^{4} + \dot{Y}^2 X Y^2 + 2 \dot{X} \dot{Y} Y^3 - M^2 X - Y^2 X &= 8 \pi \rho,  & &(\mbox{$tt$})\\
\Lambda Y^4 - 2 \ddot{Y} Y^3 - Y^2 \dot{Y}^2 + M^2 + Y^2 &=0, & &(\mbox{$xx$}) \label{Einstein^x_x}\\
\Lambda X Y^4 - \ddot{Y} X Y^3 - \ddot{X} Y^4 - \dot{X} \dot{Y} Y^3 - M^2 X &=0, & &(\mbox{$yy$})
\end{align}
with the $(zz)-$equation identical to the $(yy)-$equation and $\rho$ the density of the dust in the rest frame. The overdots mean time derivatives of the corresponding functions. Relation (\ref{Einstein^x_x}) is a second-order differential equation with two integration constants and it fixes the function $Y(\proper_time_coord)$ implicitly by the integral relation
\begin{equation}\label{integral_form_of_t}
   \proper_time_coord = \int_K^{Y(\proper_time_coord)}  \frac{\xi}{\sqrt{\tfrac{1}{3} \Lambda \xi^4 + \xi^2 + \alpha \xi - M^2}} \; \mathrm{d}\xi = \int_K^\eta  \frac{\xi}{\sqrt{f(\xi)}} \; \mathrm{d}\xi,
\end{equation}
where we use the short-hand notation $\eta:=Y(\proper_time_coord)$ and $f(\xi)$ is a master function determining the solution:
\begin{equation}\label{master_function}
  f(\xi) := \tfrac{1}{3} \Lambda \xi^4 + \xi^2 + \alpha \xi - M^2.
\end{equation}
Here, $\alpha$ is an integration constant with the dimension of length. Another integration constant is the additive constant determined by $K$, the lower limit of the integral (\ref{integral_form_of_t}), and we can remove it by shifting the time $\proper_time_coord$. We can choose $K$ arbitrarily within the range where the master function is positive. The relation (\ref{integral_form_of_t}) is given as equation (4.12) in \cite{Stewart+Ellis}. The authors of \cite{Stewart+Ellis} denote our `$\alpha$' as `$c$' and our `$M$' as `$f$', and they only give solutions for special values of the parameters appearing in the master function. In this paper, instead of integrating (\ref{integral_form_of_t}) which is complicated in general, we use it to let $\eta$ be the new time coordinate by the relation
\begin{equation}\label{new_time_coordinate}
  \mathrm{d}\proper_time_coord = \frac{\eta}{\sqrt{f(\eta)}} \mathrm{d}\eta.
\end{equation}
The consequences of zeros in the master function $f(\eta)$ will be discussed in Section (\ref{properties_of_the_integral}). So far we have
\begin{equation}
\mathrm{d}s^2 = - \frac{\eta^2}{f(\eta)}  \mathrm{d}\eta^2 + \bar{X}(\eta)^2 \mathrm{d}x^2 + \eta^2 \left( \mathrm{d}\first_coord^2 + \sinh^2 \! \first_coord \; \mathrm{d}\second_coord^2 \right),
\end{equation}
with $\bar{X}(\eta)$ replacing $X(\proper_time_coord)$. The $\first_coord\first_coord$- and $\second_coord\second_coord$-components of the Einstein equations are identical and require
\begin{equation}\label{function_X_in_terms_of_t}
  \bar{X}(\eta) = \frac{\sqrt{f(\eta)}}{\eta} \; \gamma \int_{\eta_0}^\eta \frac{\xi^2}{f(\xi)^\frac{3}{2}} \; \mathrm{d}\xi,
\end{equation}
where $\gamma$ is a dimensionless integration constant that we remove by rescaling $x$. The other integration constant is the additive constant determined by ${\eta_0}$, the lower limit of integration in (\ref{function_X_in_terms_of_t}). Therefore, ${\eta_0}$ determines $\beta$, the asymptotic value of $\bar{X}(\eta)$, as discussed in Section (\ref{properties_of_the_integral}) and also the point where $\bar{X}(\eta)$ vanishes. We thus finally have
\begin{equation}\label{the_metric}
\mathrm{d}s^2 = - \frac{\eta^2}{f(\eta)} \mathrm{d}\eta^2 + \frac{f(\eta)}{\eta^2} \left( \int_{\eta_0}^\eta \frac{\xi^2}{f(\xi)^\frac{3}{2}} \; \mathrm{d}\xi \right)^2 \mathrm{d}x^2 + \eta^2 \left( \mathrm{d}\first_coord^2 + \sinh^2 \! \first_coord \; \mathrm{d}\second_coord^2 \right),
\end{equation}
with the electromagnetic field (\ref{electromagnetic_field}) unchanged. If $f(\xi)<0$ then the factor $f(\xi)^{3/2}$ in the integral above is purely imaginary. This results in a generally complex value of the integral and does not lead to a physically plausible metric. We can certainly choose an additive constant in such a way as to cancel the purely imaginary integral, but we can only do so at a single point (we can also produce a purely imaginary integral but that results in the wrong signature of the metric). Therefore, the admissible range of $\eta$ is restricted to $f(\eta) > 0$.

Let us now assume for the moment that $\alpha \not= 0$ in (\ref{master_function}) (the special case $\alpha = 0$ will be treated below) and rescale the time coordinate $\dimensionless_time_coord := \eta / \alpha $ to obtain
\begin{equation}\label{rescaled_master_function}
f(\eta) = \tfrac{1}{3} \Lambda \eta^4 + \eta^2 + \alpha \eta - M^2 = \alpha^2 \rescaled_master_function(\dimensionless_time_coord),
\end{equation}
where
\begin{equation}
\rescaled_master_function(\dimensionless_time_coord) := \tfrac{1}{3} \Lambda \alpha^2 \dimensionless_time_coord^4 + \dimensionless_time_coord^2 + \dimensionless_time_coord - \left( \frac{M}{\alpha} \right)^{\!\!2}.
\end{equation}
The redefined master function $\rescaled_master_function(\dimensionless_time_coord)$ thus only involves two dimensionless combinations of constants, which we will now use instead of the original ones
\begin{equation}\label{rescaled_constants}
\lambda := \tfrac{1}{3} \Lambda \alpha^2,\:\: m := \frac{M}{\alpha},
\end{equation}
so that $\lambda \geq 0$ and we can write
\begin{equation}\label{rescaled_master_function_in_terms_of_rescaled_constants}
\rescaled_master_function(\dimensionless_time_coord) = \lambda \dimensionless_time_coord^4 + \dimensionless_time_coord^2 + \dimensionless_time_coord - m^2.
\end{equation}
Plugging this into the metric (\ref{the_metric}) and rescaling $x = \alpha \chi$, we can write
\begin{equation}\label{rescaled_metric}
\mathrm{d}s^2 = \alpha^2 \left[ - \frac{\dimensionless_time_coord^2}{\rescaled_master_function(\dimensionless_time_coord)} \mathrm{d}\dimensionless_time_coord^2 + \frac{\rescaled_master_function(\dimensionless_time_coord)}{\dimensionless_time_coord^2} \left( \int_\elle^\dimensionless_time_coord \frac{\zeta^2}{\rescaled_master_function(\zeta)^{\frac{3}{2}}} \; \mathrm{d}\zeta\right)^2 \mathrm{d}\chi^2 + \dimensionless_time_coord^2 \left( \mathrm{d}\first_coord^2 + \sinh^2 \! \first_coord \; \mathrm{d}\second_coord^2 \right) \right],
\end{equation}
with $\elle={\eta_0}/\alpha$. All the coordinates are dimensionless; $\alpha$ has the dimension of length and represents a constant conformal factor setting the scale of all spacetime intervals. Any scalar quantity obtained solely from the metric tensor (\ref{rescaled_metric}) and expressed in two metrics from this class differing only by $\alpha$ is thus equal up to the multiplication by a constant factor---a power of $\alpha$. The metric (\ref{rescaled_metric}) satisfies the Einstein equations in the form
\begin{equation}\label{rescaled_Einstein_equations}
G_{\mu\nu} + \frac{3}{\alpha^2}\lambda g_{\mu\nu} = 8\pi \left[ \stress_energy_tensor_{\mathrm{\scriptstyle(elmag)}{\mu\nu}} + \stress_energy_tensor_{\mathrm{\scriptstyle(dust)}{\mu\nu}} \right]
\end{equation}
with the electromagnetic field
\begin{equation}\label{rescaled_electromagnetic_field}
  F = \alpha m \sinh \first_coord \:\; \mathrm{d}\first_coord \! \wedge \! \mathrm{d}\second_coord, \;\;\; F^{ \alpha \beta} F_{ \alpha \beta} = \frac{2 m^2}{\alpha^2 \dimensionless_time_coord^4},
\end{equation}
obeying the sourceless Maxwell equations. The energy densities of the dust, $\rho$, and of the magnetic field, $\rho_{\mathrm{B}}$, as observed by comoving observers are
\begin{equation}\label{rescaled_density}
  \rho(\dimensionless_time_coord) = \frac{1}{8\pi}\frac{2}{\alpha^2 \dimensionless_time_coord \sqrt{\rescaled_master_function (\dimensionless_time_coord)} \int_\elle^\dimensionless_time_coord \zeta^2\rescaled_master_function(\zeta)^{-\frac{3}{2}} \mathrm{d}\zeta},  \;\;\; \rho_{\mathrm{B}}(\dimensionless_time_coord) = \frac{1}{16 \pi} F^{ \alpha \beta} F_{ \alpha \beta} = \frac{m^2}{8 \pi \alpha^2 \dimensionless_time_coord^4},
\end{equation}
both diverging at $\dimensionless_time_coord=0$. However, the density of the dust also diverges where the integral appearing in (\ref{rescaled_metric}) vanishes. This is due to the fact that the dust and the magnetic field only couple indirectly, through their gravity, which is not enough to force a divergence in the magnetic field.

Furthermore, the trace of the Einstein equations together with the fact that the stress-energy tensor of the electromagnetic field is traceless implies a relation between the dust density (\ref{rescaled_density}) and the Ricci scalar
\begin{equation}\label{Ricci_and_the_density}
  R = 8\pi \rho + 4 \Lambda,
\end{equation}
so that for a vanishing cosmological constant, $\Lambda = 0$, the density of the dust is proportional to the Ricci scalar. It then follows that a divergence of the dust density implies a curvature singularity although the Maxwell invariant (\ref{rescaled_electromagnetic_field}) remains finite since this does not occur at $\dimensionless_time_coord=0$ where the Maxwell invariant diverges. Therefore, the model features a truly primordial magnetic field that is finite at the moment of the initial singularity.
\section{The physics of the solution}\label{The physics of the solution}
To check the physical meaning of our solution, we will now explore various combinations of the constants appearing in the solution. In fact, the interpretation of $M$ and $\Lambda$ is clear as these correspond to the magnetic field strength, defined by the Maxwell invariant (\ref{rescaled_electromagnetic_field}), and to the asymptotic behavior (see Section (\ref{asymptotics})), respectively. We now sift through the possible cases with one or more of the 3 constants $\Lambda, \alpha, M$ vanishing in the original solution (\ref{the_metric}).

\subsection{\texorpdfstring{$\Lambda=0, M=0, \alpha=0$}{Lambda=0, M=0, alpha=0}}
This is the simplest possible case yielding $f(\eta) = \eta^2$. As $\alpha=0$, we cannot use the rescaling technique leading to (\ref{rescaled_metric}) and thus we stick to the original coordinates and metric (\ref{the_metric}). Evaluating the corresponding integral (\ref{function_X_in_terms_of_t}) explicitly and rescaling the $x$ coordinate, the metric then reads
\begin{equation}\label{the_metric_everything_zero}
\mathrm{d}s^2 = - \mathrm{d}\eta^2 + \left( \log \frac{\eta}{\eta_0} \right)^2 \mathrm{d}x^2 + \eta^2 \left( \mathrm{d}\first_coord^2 + \sinh^2 \! \first_coord \; \mathrm{d}\second_coord^2 \right),
\end{equation}
where $\eta_0$ is an integration constant. The coordinate $\eta$ measures the proper time of comoving observers, who move along time-like geodesics of the metric. This homogeneous but anisotropic spacetime is generated by pressureless dust with a density
\begin{equation}\label{density_everything_zero}
\rho = \frac{2}{8 \pi \eta^2 \log \left( \eta / \eta_0 \right)} = \frac{R}{8 \pi}.
\end{equation}
The constant $\eta_0$ thus determines the moment of the initial singularity with divergent density and Ricci scalar. It also fixes the steepness of the singularity---the rate of change of the dust density as a function of $t := \eta - \eta_0$, the proper time of comoving observers elapsed from the singularity: $\rho \sim 1/4 \pi \eta_0 t$, see Section \ref{near_the_singularity}. Ultimately, at late proper times the dust density (\ref{density_everything_zero}) dissolves as $\rho \sim 1 / 4 \pi t^2 \log t$, see Section \ref{fading_of_the_dust_density}, so that $\eta_0$ only appears in higher-order terms. Let us now redefine $\eta$ and $x$ to use new dimensionless coordinates: $\dimensionless_time_coord:= \eta/\eta_0$ and $\nu := x/\eta_0$. This yields the metric in the conformal form
\begin{equation}\label{conformal_metric_everything_zero}
\mathrm{d}s^2 = \eta_0^2 \left[ - \mathrm{d}\dimensionless_time_coord^2 + \left( \log \dimensionless_time_coord\right)^2 \mathrm{d}\nu^2 + \dimensionless_time_coord^2 \left( \mathrm{d}\first_coord^2 + \sinh^2 \! \first_coord \; \mathrm{d}\second_coord^2 \right) \right].
\end{equation}
All the coordinates are dimensionless; $\eta_0$ has the dimension of length and represents a constant conformal factor setting the scale of all spacetime intervals. We can rescale the unit of length we use to measure the distances, getting thus rid of $\eta_0$ altogether.
\subsection{\texorpdfstring{$\Lambda = 0, M = 0, \alpha \not= 0$}{Lambda = 0, M = 0, alpha not= 0}}\label{Lambda = 0, M = 0, alpha not= 0}
\begin{figure}[h]
\centering
\includegraphics[width = 0.7\textwidth]{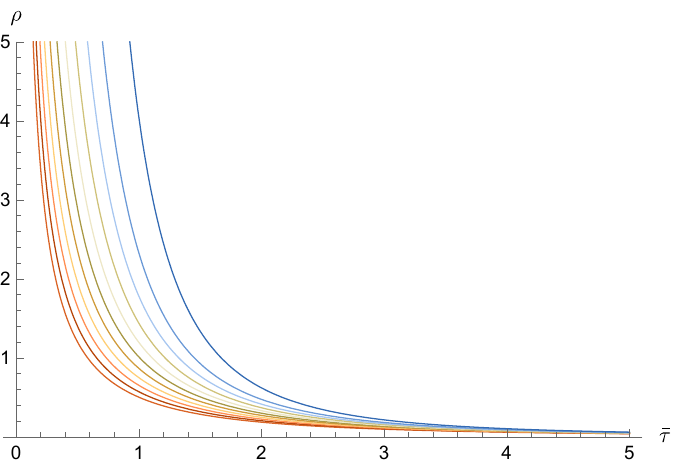}
\caption{The density of the dust with $\Lambda = 0, M = 0,$ and $8\pi\alpha^2 = 1$. We plot the density curves for $\beta = \{-1, -0.9, -0.8, \ldots, 0\}$ in (\ref{integral_Lambda=0_M=0_alpha_non_zero}), from orange to blue. The density diverges at a certain time $\dimensionless_time_coord_0>0$ determined by $\beta$ and it is positive at all times afterwards, vanishing asymptotically. We use a shifted time coordinate $\bar{\dimensionless_time_coord} := \dimensionless_time_coord - \dimensionless_time_coord_0$ so that the singularity always occurs at $\bar{\dimensionless_time_coord}=0$.} \label{plot_density_only_alpha}
\end{figure}

We obtain an exact dust solution with the rescaled master function (\ref{rescaled_master_function_in_terms_of_rescaled_constants})
\begin{equation}
  \rescaled_master_function(\dimensionless_time_coord) = \dimensionless_time_coord(\dimensionless_time_coord+1).
\end{equation}
The rescaled metric (\ref{rescaled_metric}) can now be evaluated explicitly, since
\begin{equation}\label{integral_Lambda=0_M=0_alpha_non_zero}
\int_\elle^\dimensionless_time_coord \frac{\zeta^2 \; \mathrm{d}\zeta}{\left[\zeta(\zeta+1)\right]^{\frac{3}{2}}} = \int_\elle^\dimensionless_time_coord \frac{\zeta^{\frac{1}{2}} \; \mathrm{d}\zeta}{\left(\zeta+1\right)^{\frac{3}{2}}} = 2 \left( \mathrm{arctanh} \sqrt{\frac{\dimensionless_time_coord}{\dimensionless_time_coord+1}} - \sqrt{\frac{\dimensionless_time_coord}{\dimensionless_time_coord+1}} \; \right) + \beta,
\end{equation}
with an integration constant $\beta = - 2 \left( \mathrm{arctanh} \sqrt{\frac{\elle}{\elle+1}} - \sqrt{\frac{\elle}{\elle+1}} \; \right) \leq 0$ and the lower limit of integration $\elle \geq 0$ (to avoid complex values in the integrand) so that at $\dimensionless_time_coord=\elle$, the integral vanishes. We can now write the metric explicitly as
\begin{equation}\label{rescaled_Lambda=0_M=0_alpha_non_zero}
  \mathrm{d}s^2 \! = \! \alpha^2 \! \left\{ \! - \frac{\dimensionless_time_coord}{\dimensionless_time_coord+1} \mathrm{d}\dimensionless_time_coord^2 + 4 \left[ \sqrt{\frac{\dimensionless_time_coord+1}{\dimensionless_time_coord}} \left( \mathrm{arctanh} \sqrt{\frac{\dimensionless_time_coord}{\dimensionless_time_coord+1}} + \frac{\beta}{2} \right) - 1 \right]^2 \!\!\! \mathrm{d}\chi^2 + \dimensionless_time_coord^2 \left( \mathrm{d}\first_coord^2 + \sinh^2 \! \first_coord \; \mathrm{d}\second_coord^2 \right) \right\}.
\end{equation}
We can also use the addition formula for arctanh to put the expression occurring in the line element in the following form
\begin{equation}
  \mathrm{arctanh} \sqrt{\frac{\dimensionless_time_coord}{\dimensionless_time_coord+1}} + \frac{\beta}{2} = \mathrm{arctanh} \left( \frac{\sqrt{\dimensionless_time_coord (\dimensionless_time_coord+1)} - \sqrt{\elle (\elle +1)}}{\dimensionless_time_coord + \elle +1} \right) + \sqrt{\frac{\elle}{\elle + 1}}.
\end{equation}
The density is proportional to the Ricci scalar (see (\ref{Ricci_and_the_density})) and behaves as
\begin{equation}\label{density_only_alpha}
  \rho = \frac{1}{8\pi\alpha^2} \frac{1}{\dimensionless_time_coord^2 \left(\sqrt{\frac{\dimensionless_time_coord+1}{\dimensionless_time_coord}} \; \left( \mathrm{arctanh} \sqrt{\frac{\dimensionless_time_coord}{\dimensionless_time_coord+1}} + \frac{\beta}{2} \right) - 1 \right)} = \frac{R}{8\pi}.
\end{equation}
For $\beta = 0$, there is a singularity at $\dimensionless_time_coord=0$, where the denominator vanishes, and $\rho \sim 3/8 \pi \alpha^2 \dimensionless_time_coord^3$ or in terms of the proper time of comoving observers elapsed from the singularity $\rho \sim 1/6 \pi t^2$. This is not enough to determine the value of $\alpha$, so we need to go one order higher in the expansion, yielding
\begin{equation}
\rho \sim \frac{1}{6 \pi t^2} - \frac{1}{5\pi (96 \alpha^2 t^4)^{1/3}}.
\end{equation}
By following the evolution of the dust density as a function of their proper time shortly after the initial singularity, comoving observers can thus determine the value of $\alpha$. Due to the fact that, as a function of $\dimensionless_time_coord$, the integral (\ref{integral_Lambda=0_M=0_alpha_non_zero}) with $\beta=0$ is a continuous, increasing function that ranges over the entire non-negative real axis, choosing $\beta<0$ in (\ref{integral_Lambda=0_M=0_alpha_non_zero}) shifts the singularity in time and controls its steepness but cannot remove it, see Figure (\ref{plot_density_only_alpha}). Taylor expanding the dust density near the singularity, we obtain
\begin{equation}
\rho \sim \frac{1}{4 \pi} \frac{\sqrt{\elle (1 + \elle) }}{\alpha \elle^2} \frac{1}{t} - \frac{3}{8 \pi} \frac{\elle (1 + \elle)}{\alpha ^2 \elle^4} + \frac{1}{48 \pi} \frac{\sqrt{\elle (1+\elle)}(27 + 23 \elle)}{\alpha^3 \elle^5} t,
\end{equation}
which again enables us to determine the values of both $\alpha$ and $\elle$ (the first two terms are not sufficient since they contain the same combination of $\alpha$ and $\elle$).

The solution (\ref{rescaled_Lambda=0_M=0_alpha_non_zero}) even works asymptotically since for $\dimensionless_time_coord \gg 1$ we can write the metric as
$$\mathrm{d}s^2 = \alpha^2 \left\{ - \mathrm{d}\dimensionless_time_coord^2 + \left( \log \dimensionless_time_coord \right)^2 \mathrm{d}\chi^2 + \dimensionless_time_coord^2 \left( \mathrm{d}\first_coord^2 + \sinh^2 \! \first_coord \; \mathrm{d}\second_coord^2 \right) \right\},$$
which is identical to the solution (\ref{conformal_metric_everything_zero}) above, with $\alpha$ playing the role of $\eta_0$ now. We thus get the same late time behavior of the dust density in terms of the proper time $\rho \sim 1 / 4 \pi t^2 \log t$.
\subsection{\texorpdfstring{$\Lambda = 0, M \not= 0, \alpha \not= 0$}{Lambda = 0, M not= 0, alpha not= 0}}\label{Lambda = 0, M not= 0, alpha not= 0}
\begin{figure}[h]
\centering
\includegraphics[width = 0.7\textwidth]{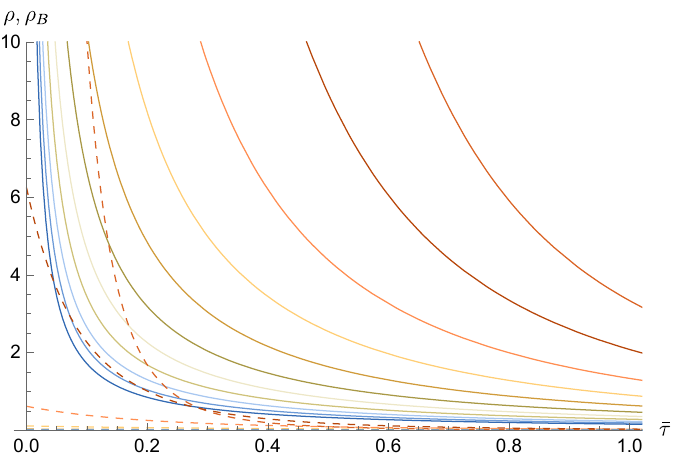}
\caption{The energy density of the dust, $\rho$ (\ref{Density_only_alpha_and_M}), (solid lines) and of the magnetic field, $\rho_{\mathrm{B}}$ (\ref{rescaled_density}), (dashed lines) with $8\pi\alpha^2=1,$ and $m = \{ 0.1, 0.3, 0.5, \ldots, 2.1 \}$, from orange to blue. The dust density diverges at a certain time $\dimensionless_time_coord_0>0$ determined by $m$ and is positive at all times afterwards. For the dependence of $\dimensionless_time_coord_0$ on the integration constant $\beta$ in (\ref{the_integral_only_alpha_and_M}), see Figure (\ref{diverging_density}). We use a shifted time coordinate $\bar{\dimensionless_time_coord} := \dimensionless_time_coord - \dimensionless_time_coord_0$ so that the singularity always occurs at $\bar{\dimensionless_time_coord}=0$.} \label{plot_density_only_alpha_and_M}
\end{figure}

\begin{figure}[h]
\centering
\includegraphics[width = 0.7\textwidth]{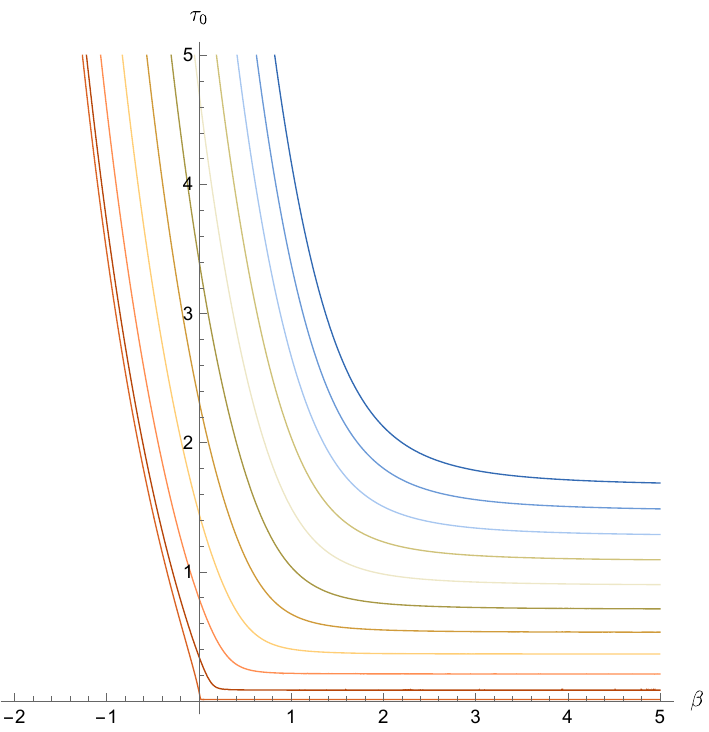}
\caption{The moment of singularity---diverging dust density (\ref{Density_only_alpha_and_M})---$\dimensionless_time_coord_0$, as a function of the integration constant $\beta$ chosen in the integral (\ref{the_integral_only_alpha_and_M}) with $m = \{ 0.1, 0.3, 0.5, \ldots, 2.1 \}$, from orange to blue. The asymptotic values of the curves as $\beta \rightarrow \infty$ are given by the vanishing of the term $\sqrt{\rescaled_master_function(\dimensionless_time_coord)} = \sqrt{\dimensionless_time_coord^2 + \dimensionless_time_coord - m^2}$ appearing in the denominator of (\ref{rescaled_density}).} \label{diverging_density}
\end{figure}

As a slight generalization of the previous case, we obtain an exact dust plus magnetic field solution with the master function
\begin{equation}
  \rescaled_master_function(\dimensionless_time_coord) = \dimensionless_time_coord^2 + \dimensionless_time_coord - m^2.
\end{equation}
The proper time of comoving observers is
\begin{equation}
  \frac{\proper_time_coord}{\alpha} = \int_k^\dimensionless_time_coord \frac{\zeta \: \mathrm{d}\zeta}{\sqrt{\zeta^2 + \zeta - m^2}} = \sqrt{\dimensionless_time_coord^2 + \dimensionless_time_coord - m^2}+\frac{1}{2} \log \left( 1 + 2 \dimensionless_time_coord - 2 \sqrt{\dimensionless_time_coord^2 + \dimensionless_time_coord - m^2} \right),
\end{equation}
with $k = K/\alpha$ and the additive constant removed by shifting $t$. The required integral in the metric (\ref{rescaled_metric}) reads
\begin{equation}\label{the_integral_only_alpha_and_M}
\int_\elle^\dimensionless_time_coord \frac{\zeta^2 \; \mathrm{d}\zeta}{\left( \zeta^2 + \zeta - m^2 \right)^{\frac{3}{2}}} = 2 \frac{m^2 - \dimensionless_time_coord ( 1+ 2m^2 )}{\left(1+4 m^2\right) \sqrt{\dimensionless_time_coord^2 + \dimensionless_time_coord - m^2}}-\log \left( 1 + 2 \dimensionless_time_coord - 2 \sqrt{\dimensionless_time_coord^2 + \dimensionless_time_coord - m^2} \right) + \beta,
\end{equation}
with $\beta$ an additive integration constant determined by $\elle$. Regardless of $\beta$, the integral ranges over the entire real axis with a single zero so we first choose $\beta=0$. The metric then reads
\begin{eqnarray}
    \mathrm{d}s^2 &=& \alpha^2 \left\{ - \frac{\dimensionless_time_coord^2}{\dimensionless_time_coord^2 + \dimensionless_time_coord - m^2} \mathrm{d}\dimensionless_time_coord^2 + \dimensionless_time_coord^2 \left( \mathrm{d}\first_coord^2 + \sinh^2 \! \first_coord \; \mathrm{d}\second_coord^2 \right) + \frac{\dimensionless_time_coord^2 + \dimensionless_time_coord - m^2}{\dimensionless_time_coord^2} \cdot \right. \nonumber\\
    && \left. \cdot \left[ 2 \frac{m^2 - \dimensionless_time_coord ( 1+ 2m^2 )}{\left(1+4 m^2\right) \sqrt{\dimensionless_time_coord^2 + \dimensionless_time_coord - m^2}}-\log \left( 1 + 2 \dimensionless_time_coord - 2 \sqrt{\dimensionless_time_coord^2 + \dimensionless_time_coord - m^2} \right) \right]^2 \mathrm{d}\chi^2 \right\}.
\end{eqnarray}
The density of the dust is proportional to the Ricci scalar (see (\ref{Ricci_and_the_density})):
\begin{equation}\label{Density_only_alpha_and_M}
  \rho = \frac{R}{8\pi} = \frac{1}{8\pi}\frac{2}{\alpha^2 \dimensionless_time_coord \left[ 2 \frac{m^2 - \dimensionless_time_coord ( 1+ 2m^2 )}{1+4 m^2} - \sqrt{\dimensionless_time_coord^2 + \dimensionless_time_coord - m^2} \log \left( 1 + 2 \dimensionless_time_coord - 2 \sqrt{\dimensionless_time_coord^2 + \dimensionless_time_coord - m^2} \right) \right]},
\end{equation}
see Figure (\ref{plot_density_only_alpha_and_M}). Asymptotically, this is the same as the previous solution (\ref{rescaled_Lambda=0_M=0_alpha_non_zero}) and (\ref{density_only_alpha}) with no magnetic field. Choosing $\beta \not= 0$ in the integral (\ref{the_integral_only_alpha_and_M}) shifts the moment of singularity in time and changes its slope but cannot remove it altogether, see Figure (\ref{diverging_density}).

\begin{figure}[h]
\centering
\subfloat[We fix $\beta=0$ in (\ref{the_integral_only_alpha_and_M}) and, from orange to blue, we set $m = \{ 0.1, 0.3, 0.5, \ldots, 2.1 \}$ in (\ref{Density_only_alpha_and_M}). We can see that, in this case, the influence of the magnetic field on the evolution of the universe is limited, since the highest value of the ratio is about 2.3\% for $m \approx 0.308$.]{\includegraphics[width = 0.45\textwidth]{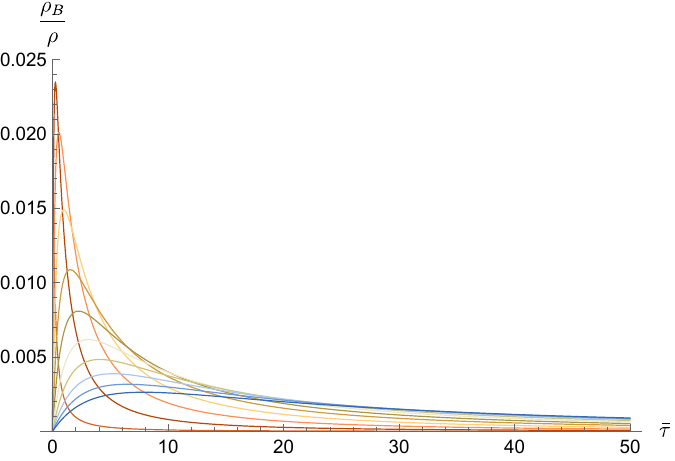}\label{magnetic_to_dust_energy_density_ratio_zero_beta}}
\hspace{1em}
\subfloat[We fix $m=0.13$ and, orange to blue, we use $\beta = \{0.1, 0.2, 0.3, \ldots, 1\}$ in (\ref{the_integral_only_alpha_and_M}), changing (\ref{Density_only_alpha_and_M}) accordingly. Now the magnetic field dominates the dust in a certain era after the initial singularity.]{\includegraphics[width = 0.45\textwidth]{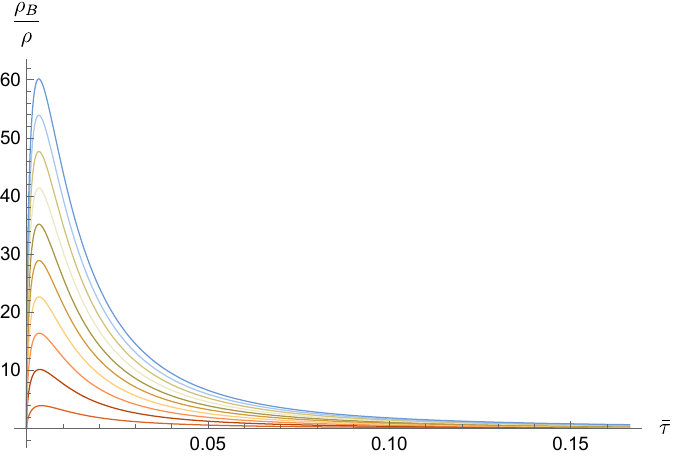}\label{magnetic_to_dust_energy_density_ratio_non_zero_beta}}
\caption{The ratio of the energy density of the magnetic field $\rho_{\mathrm{B}}$ (\ref{rescaled_density}) to that of the dust $\rho$ (\ref{Density_only_alpha_and_M}) for $\Lambda=0$. In these plots, we use a new time coordinate $\bar{\dimensionless_time_coord} := \dimensionless_time_coord - \dimensionless_time_coord_0$ so that the singularity always occurs at $\bar{\dimensionless_time_coord}=0$. Both at the singularity and in the asymptotic region $\bar{\dimensionless_time_coord} \rightarrow \infty$, the energy of the magnetic field is always negligible compared to that of the dust.}
\end{figure}

We can also compare the influence of the dust and of the magnetic field on the gravitational field by calculating the ratio of their energy densities
\begin{equation}\label{energy_density_ratio}
  \frac{\rho_{\mathrm{B}}}{\rho} = \frac{m^2 \sqrt{\rescaled_master_function(\dimensionless_time_coord)} }{2 \dimensionless_time_coord^3} \int_\elle^\dimensionless_time_coord \zeta^2 \rescaled_master_function(\zeta)^{-\frac{3}{2}} \mathrm{d}\zeta.
\end{equation}
The result is plotted in Figure (\ref{magnetic_to_dust_energy_density_ratio_zero_beta}) for the integral (\ref{the_integral_only_alpha_and_M}) with $\beta=0$, and in Figure (\ref{magnetic_to_dust_energy_density_ratio_non_zero_beta}) with $\beta \not= 0$. In the latter case, the ratio can exceed 1 for a certain period of time while it always approaches 0 at the singularity and for large times.
\subsection{\texorpdfstring{$\Lambda \not= 0, M \not=0, \alpha = 0$}{Lambda not= 0, M not=0, alpha = 0}}\label{alpha_vanishing}
Since $\alpha$ is zero, we need to go back to the metric of the form (\ref{the_metric}) and the master function can now be expressed in the following form
\begin{equation}\label{alpha=0}
  f(\eta) = \frac{\Lambda}{3} \left( \eta^2 + a^2 \right) \left( \eta^2 - b^2 \right),
\end{equation}
with
\begin{equation}
  a^2 = \frac{3}{2 \Lambda } \left( 1 + \sqrt{ 1 + \frac{4}{3} \Lambda  M^2} \right), \;\; b^2 = \frac{3}{2 \Lambda } \left(- 1 + \sqrt{ 1 + \frac{4}{3} \Lambda  M^2} \right),
\end{equation}
which is physically plausible for $\Lambda>0$. We further find
\begin{equation}\label{invariants_just_alpha}
  F^{ \alpha \beta} F_{ \alpha \beta} = \frac{2 M^2}{\eta^4}, \;\;\; \rho(\eta) = \frac{1}{8\pi}\frac{2}{\eta \sqrt{f(\eta)} \int_{\eta_0}^\eta \xi^2 f(\xi)^{-\frac{3}{2}} \mathrm{d}\xi      }.
\end{equation}
Let us now rewrite the integral appearing in the density above (we take out the $\Lambda$ factor for brevity)
\begin{eqnarray}
   \left( \frac{\Lambda}{3} \right)^{\!\! \frac{3}{2}} \!\!\! \int_{\eta_0}^\eta \! \frac{\xi^2}{f(\xi)^\frac{3}{2}} \; \mathrm{d}\xi &=& \int_{\eta_0}^\eta \! \frac{\xi^2}{\left( (\xi^2 + a^2) (\xi^2 - b^2) \right)^\frac{3}{2}} \; \mathrm{d}\xi = \int_{\eta_0}^\eta \! \frac{1}{4\xi}\frac{4\xi \left( \xi^2 + \frac{a^2 - b^2}{2}\right) - 4\xi \frac{a^2 - b^2}{2}}{\left( (\xi^2 + a^2) (\xi^2 - b^2) \right)^\frac{3}{2}} \; \mathrm{d}\xi = \nonumber\\
   &=& - \frac{1}{2\xi} \frac{1}{\left( (\xi^2 + a^2) (\xi^2 - b^2) \right)^\frac{1}{2}} - \frac{1}{2} \int_{\eta_0}^\eta \! \frac{\mathrm{d}\xi}{\xi^2\left( (\xi^2 + a^2) (\xi^2 - b^2) \right)^\frac{1}{2}} \nonumber\\
   && - \frac{a^2 - b^2}{2} \int_{\eta_0}^\eta \! \frac{\mathrm{d}\xi}{\left( (\xi^2 + a^2) (\xi^2 - b^2) \right)^\frac{3}{2}}.\label{partial_integrals}
\end{eqnarray}
Let us further note that, if all the integrals exist and are finite (see Section (\ref{properties_of_the_integral})), we can write
\begin{eqnarray*}
\int_{{\eta_0}}^\eta h(\xi) \mathrm{d}\xi &=& \int_{{\eta_0}}^\infty h(\xi) \mathrm{d}\xi + \int_\infty^\eta h(\xi) \mathrm{d}\xi = \int_{{\eta_0}}^\infty h(\xi) \mathrm{d}\xi - \int_\eta^\infty h(\xi) \mathrm{d}\xi = \nonumber\\
&=& \mathrm{const.}  - \int_\eta^\infty h(\xi) \mathrm{d}\xi.
\end{eqnarray*}
Denoting the above constant as $\beta$ enables us to rewrite the sought integral as
$$\int_{\eta_0}^\eta \frac{\xi^2}{f(\xi)^\frac{3}{2}} \; \mathrm{d}\xi = \beta - \int_\eta^\infty  \frac{\xi^2}{f(\xi)^\frac{3}{2}} \; \mathrm{d}\xi.$$
Gradshteyn \& Ryzhik (3.156.4) \cite{Gradshteyn_Ryzhik_2007} define $\chi := \arcsin \sqrt{\frac{{a^2 + b^2}}{{a^2 + \eta^2}}}, s := \frac{a}{\sqrt{a^2 + b^2}}$ and state
$$\int_\eta^\infty \frac{\mathrm{d}\xi}{\xi^2\left( (\xi^2 + a^2) (\xi^2 - b^2) \right)^\frac{1}{2}} = \frac{\left( a^2 + b^2 \right) E(\chi;s) - b^2 F(\chi;s)}{a^2 b^2 \sqrt{a^2 + b^2}} - \frac{1}{b^2 \eta} \sqrt{\frac{\eta^2 - b^2}{a^2 + \eta^2}},$$
where $\eta \geq b > 0$ and $F(\chi;s) = \int_{0}^{\chi} \mathrm{d}a/\sqrt{1-s^2 \sin^2 a}$ and $E(\chi;s) = \int_{0}^{\chi} \sqrt{1-s^2 \sin^2 a} \; \mathrm{d}a$ are elliptic integrals of the 1\textsuperscript{st} and 2\textsuperscript{nd} kind, respectively.\footnote{There is an error in (3.156.4) in the 8th edition of Gradshteyn and Ryzhik \cite{Gradshteyn_Ryzhik_2015}: it defines $\chi := \arcsin \frac{\eta}{b} \sqrt{\frac{{a^2 + b^2}}{{a^2 + \eta^2}}}$, which has the wrong asymptotics for $\eta \rightarrow \infty$. We confirmed the result from the 7th edition \cite{Gradshteyn_Ryzhik_2007} by integrating numerically.} Gradshteyn \& Ryzhik (3.163.4) further state
\begin{eqnarray*}
\int_\eta^\infty \frac{\mathrm{d}\xi}{\left( (\xi^2 + a^2) (\xi^2 - b^2) \right)^\frac{3}{2}} &=& \frac{\left( b^2 - a^2 \right) E(\chi;s) - b^2 F(\chi;s)}{a^2 b^2 \left( a^2 + b^2 \right)^{\frac{3}{2}}}\\
&& + \frac{\eta}{b^2 \left( a^2 + b^2 \right) \sqrt{\left(\eta^2 + a^2 \right) \left(\eta^2 - b^2\right)}},
\end{eqnarray*}
where $\eta > b > 0$. Plugging all this into the expanded integral (\ref{partial_integrals}), we have an explicit expression for the integral appearing in the metric (\ref{the_metric}) and density (\ref{invariants_just_alpha}):
\begin{eqnarray}
   && \left( \frac{\Lambda}{3} \right)^\frac{3}{2} \int_{\eta_0}^\eta \frac{\xi^2}{f(\xi)^\frac{3}{2}} \; \mathrm{d}\xi = - \frac{1}{2\eta} \frac{1}{\sqrt{ (\eta^2 + a^2) (\eta^2 - b^2) }} \nonumber\\
   && + \frac{1}{2} \left[ \frac{\left( a^2 + b^2 \right) E(\chi;s) - b^2 F(\chi;s)}{a^2 b^2 \sqrt{a^2 + b^2}} - \frac{1}{b^2 \eta} \sqrt{\frac{\eta^2 - b^2}{a^2 + \eta^2}} \right] \nonumber\\
   && + \frac{a^2 - b^2}{2} \left[ \frac{\left( b^2 - a^2 \right) E(\chi;s) - b^2 F(\chi;s)}{a^2 b^2 \left( a^2 + b^2 \right)^{\frac{3}{2}}} + \frac{\eta}{b^2 \left( a^2 + b^2 \right) \sqrt{\left(\eta^2 + a^2 \right) \left(\eta^2 - b^2\right)}} \right] + \beta = \nonumber\\
   && = - \frac{1}{2\eta} \frac{1}{\sqrt{ (\eta^2 + a^2) (\eta^2 - b^2) }} - \frac{1}{2 b^2 \eta} \sqrt{\frac{\eta^2 - b^2}{a^2 + \eta^2}}  + \frac{2 E(\chi;s) - F(\chi;s)}{\left( a^2 + b^2 \right)^{\frac{3}{2}}}\nonumber\\
   && + \frac{\eta \left(a^2 - b^2 \right)}{2 b^2 \left( a^2 + b^2 \right) \sqrt{\left(\eta^2 + a^2 \right) \left(\eta^2 - b^2\right)}} + \beta. \label{primitive_function_Lambda_M}
\end{eqnarray}
The asymptotic value of (\ref{primitive_function_Lambda_M}) is equal to the additive integration constant $\beta$. Choosing a particular value $\beta>0$, we can now plug (\ref{primitive_function_Lambda_M}) into the expression for the density of the dust (\ref{invariants_just_alpha}). The resulting plots look very similar to Figure (\ref{plot_density_only_alpha}) so we do not include them here. This solution is equivalent to (31a-b) of \cite{Lorenz}. For a physical interpretation of all parameters appearing in the solution see the following section.
\section{The general case}\label{The general case}
The previous example with $\Lambda \not= 0, M \not=0, \alpha = 0$ is another exact solution that we are able to write explicitly due to the fact that in this case, the master function (\ref{master_function}) reduces to a quadratic function of $\eta^2$. Generally however, the master function is a quartic polynomial. The integral in (\ref{the_metric}) can still be evaluated explicitly in terms of elliptic integrals but it becomes very complicated. Let us first explore some properties of the general solution that can be deduced without knowing its explicit form. In the following subsections \ref{asymptotics}--\ref{near_the_singularity}, we will use the original, non-rescaled, form of the metric (\ref{the_metric}) with the master function defined by (\ref{master_function}) to be able to discuss both $\alpha = 0$ and $\alpha \not= 0$ cases together.
\subsection{Asymptotic properties at large times}\label{asymptotics}
Let us first assume $\Lambda \not= 0$. After rescaling the $x$ and $\eta$ coordinates the asymptotic form of the solution at $\eta \rightarrow \infty$ and with a positive asymptotic value of the integral appearing in the metric (\ref{the_metric}) is
\begin{equation}\label{asymptotic_form_of_the_metric}
\mathrm{d}s^2 = \frac{3}{\Lambda} \left[ - \frac{\mathrm{d}\eta^2}{\eta^2}  + \eta^2 \mathrm{d}x^2 + \eta^2 \left( \mathrm{d}\first_coord^2 + \sinh^2 \! \first_coord \; \mathrm{d}\second_coord^2 \right) \right].
\end{equation}
This is not an exact solution, it only satisfies the vacuum Einstein equations asymptotically for large $\eta$. However, this is the asymptotic form of the de Sitter space (4.21) of \cite{Griffiths+Podolsky} for $\eta \rightarrow \infty$ as expected since $\Lambda$ prevails here.

If, on the other hand, the cosmological constant vanishes, $\Lambda=0$, the asymptotic form of the metric is identical to (\ref{the_metric_everything_zero}).
\subsection{The integral}\label{properties_of_the_integral}
Let us examine the integral in (\ref{the_metric}). Assume that $\Lambda \geq 0$ so that there is at least one positive root of $f(\eta)$. Further assume that the largest positive root $\eta_{f}$ is a simple root with $f(\eta) \approx \delta \cdot (\eta - \eta_{f})$ in its vicinity---we will see in section \ref{factorization} that this is the only physically plausible case. Then, necessarily, $\delta>0$. Inspecting the integral near the root, we find
\begin{equation}\label{approximate_integral}
  \int  \frac{\eta^2}{f(\eta)^\frac{3}{2}} \; \mathrm{d}\eta \sim \int  \frac{\eta^2}{\delta^\frac{3}{2}(\eta - \eta_{f})^\frac{3}{2}} \; \mathrm{d}\eta = -\frac{2\eta_{f}^2}{\delta^\frac{3}{2}(\eta - \eta_{f})^\frac{1}{2}},
\end{equation}
which diverges to negative infinity from above at the root. Asymptotically, on the other hand, we have
\begin{equation}\label{integral_asymptotic_form}
  \int  \frac{\eta^2}{f(\eta)^\frac{3}{2}} \; \mathrm{d}\eta \sim \int \frac{\eta^2}{(\frac{\Lambda}{3} \eta^4)^\frac{3}{2}} \; \mathrm{d}\eta = -\frac{\sqrt{3}}{\Lambda^\frac{3}{2}\eta^3} + \beta \rightarrow \beta,
\end{equation}
with $\beta$ an arbitrary integration constant, refer to Figure (\ref{plot_integral}). This implies that the integral has a finite asymptotic value $\beta$ determined by our choice of the additive integration constant. As the integrand is positive, the integral is an increasing function. We now choose $\beta>0$ so that there exists a single value $\eta_0 > \eta_{f}$ for which the integral vanishes while $f(\eta_0)>0$. The Ricci scalar is
\begin{equation}\label{Ricci_scalar}
  R = \frac{2}{\eta \sqrt{f(\eta)} \int_{\eta_0}^\eta \xi^2f(\xi)^{-\frac{3}{2}} \mathrm{d}\xi} + 4 \Lambda = 8\pi \rho + 4 \Lambda.
\end{equation}
It goes to $4\Lambda$ asymptotically, consistent with the dS asymptotics, and it diverges at $\eta_0$ along with the density, signifying a curvature singularity---while it is finite at $\eta=\eta_f$. This enables observers to determine the cosmological constant from the asymptotic value of the Ricci scalar since the dust density vanishes asymptotically, see the next section. The total proper time of a comoving observer is
\begin{equation}\label{proper_time}
  \proper_time_coord =  \int_{\eta_0}^\eta \frac{\xi}{\sqrt{f(\xi)}} \mathrm{d}\xi,
\end{equation}
which is zero at the moment of singularity, $\eta_0$, and yields $\proper_time_coord \sim \sqrt{\frac{3}{\Lambda}} \log \frac{\eta}{H}$ at infinity for a non-vanishing cosmological constant, with $H$ an integration constant. If $\Lambda=0$, then we have $\proper_time_coord \sim \eta$ asymptotically. The proper time thus diverges for large times as desired in both cases---the spacetime cannot be extended. It is restricted by the curvature singularity at $\eta = \eta_0>\eta_{f}$ from below and then it tends to $+\infty$.
\subsection{Energy densities---the asymptotic fading}\label{fading_of_the_dust_density}
Let us convert the dust and magnetic field densities (\ref{rescaled_density}) to the original coordinate $\eta$ to find
\begin{equation}\label{density}
  \rho(\eta) = \frac{1}{4\pi}\frac{1}{\eta \sqrt{f(\eta)} \int_{\eta_0}^\eta \xi^2 f(\xi)^{-\frac{3}{2}} \mathrm{d}\xi},  \;\;\; \rho_{\mathrm{B}}(\eta) = \frac{M^2}{8 \pi \eta^4}.
\end{equation}
Now first assume $\Lambda \not= 0$ and expand the dust density in terms of $\eta$ for $\eta \rightarrow \infty$ as in (\ref{integral_asymptotic_form}) to obtain
\begin{equation}
    \rho \sim \frac{1}{4 \pi} \, \sqrt{\frac{3}{\Lambda}} {\frac {1}{\beta {\eta}^{3}}},
\end{equation}
or in terms of the proper time $t$ of comoving observers
\begin{equation}
    \rho \sim \frac{1}{4 \pi} \, \sqrt{\frac{3}{\Lambda}} \frac{1}{\beta H^3} \exp \left( -3 \sqrt{\frac{\Lambda}{3}} t \right),
\end{equation}
with the integration constant $H$ introduced below (\ref{proper_time}), so that the asymptotic value of the integral, $\beta$ of (\ref{integral_asymptotic_form}), governs the rate of decay of the dust density for large times as the spacetime approaches the de Sitter form of (\ref{asymptotic_form_of_the_metric}). Generally, there is no simple formula for $\beta$ as a function of the constant $\eta_0$ appearing as the lower limit of integration in (\ref{density}). However, if $\eta_0 \gg \eta_f$ with $\eta_f > 0$ the root of the master function $f(\eta)$, we find
\begin{equation}
    \beta \sim \frac{\sqrt{3}}{\Lambda^\frac{3}{2}\eta_0^3},
\end{equation}
which we confirmed numerically. We thus have
\begin{equation}
    \rho \sim \frac{\Lambda }{4 \pi} \left( \frac{\eta_0}{\eta} \right)^3.
\end{equation}
With $H \sim \eta_0$ now, this expression can again be written using the asymptotic proper time of comoving observers as
\begin{equation}\label{dust_energy_asymptotics_Lambda}
    \rho \sim \frac{\Lambda}{4 \pi} \exp \left( -3 \sqrt{\frac{\Lambda}{3}} t \right).
\end{equation}

For the energy density of the magnetic field (\ref{density}), we find asymptotically
\begin{equation}\label{magnetic_field_energy_asymptotics_Lambda}
    \rho_{\mathrm{B}} \sim \frac{M^2}{8 \pi H^4} \exp \left( -4 \sqrt{\frac{\Lambda}{3}}\ t \right),
\end{equation}
with $H \sim \eta_0$ again if $\eta_0 \gg \eta_f$. Unfortunately, due to the presence of $H$, these relations do not allow us to determine the 3 remaining parameters of the spacetime. If, on the other hand, the cosmological constant vanishes, then as the proper time $t$ of comoving observers $t \rightarrow \infty$, we find
\begin{equation}\label{dust_energy_asymptotics}
    \rho \sim \frac{1}{4 \pi} \, \frac {1}{t^2 \log t},
\end{equation}
while the energy density of the magnetic field yields
\begin{equation}\label{magnetic_field_energy_asymptotics}
    \rho_{\mathrm{B}} \sim \frac{M^2}{8 \pi} \frac{1}{t^4}.
\end{equation}
To summarize, irregardless of the cosmological constant, we have $\rho>0$ for $\eta \in (\eta_0,\infty)$ so that the energy conditions are satisfied and $g_{xx}>0, g_{tt}<0$ as required for the correct metric signature. The system starts with a singularity and then it expands with the density of dust and density of electromagnetic energy fading away. If $\Lambda=0$ then the comoving observers are able to ascertain the value of the constant $M$ from the asymptotic rate of fading of the magnetic field energy (\ref{magnetic_field_energy_asymptotics}).
\subsection{Energy densities near the singularity}\label{near_the_singularity}
\begin{figure}[h]
\centering
\subfloat[{The case $\Lambda = 0, M \not= 0, \alpha \not= 0$ of $\rho$ (\ref{Density_only_alpha_and_M}) with $8\pi\alpha^2=1$ and $m = \{ 0.1, 0.3, 0.5, \ldots, 2.1 \}$, from orange to blue, as in \autoref{plot_density_only_alpha_and_M}.}]{\includegraphics[width = 0.45\textwidth]{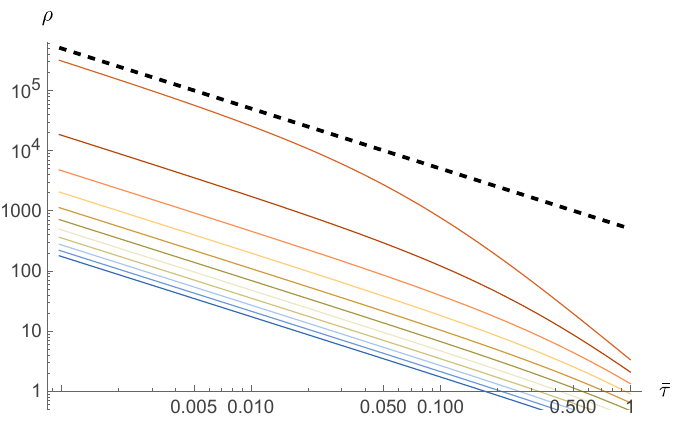}
}
\hspace{1em}
\subfloat[{The general case of $\rho$ (\ref{rescaled_density}) with $\lambda = 1, \beta \approx 0.18$ and $m = \{0.1, 0.3, 0.5, \ldots, 2.1\}$, from orange to blue, as in \autoref{dust_density_non_zero_beta_non_zero_lambda}.}]{\includegraphics[width = 0.45\textwidth]{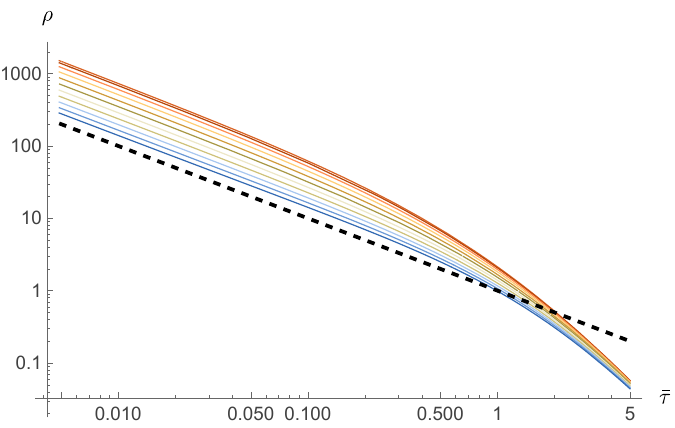}
}
\caption{The energy density of the dust $\rho$ near the singularity at $\dimensionless_time_coord = \elle$, plotted in rescaled variables. Both axes are scaled logarithmically to capture the asymptotic behavior of the density near the singularity. We use a shifted time coordinate $\bar{\dimensionless_time_coord} := \dimensionless_time_coord - \dimensionless_time_coord_0$ so that the singularity always occurs at $\bar{\dimensionless_time_coord}=0$. The dotted black reference lines are $\rho \sim 1/\bar{\dimensionless_time_coord}$. We can see that the density indeed follows the relation (\ref{log_log_density_dependence}) as the curves approach straight lines parallel to the reference line.}
\label{log_log_plot_density_only_alpha_and_M}
\end{figure}

\begin{figure}[h]

\centering
\subfloat[{The red curve is the theoretical value of $W$, the crosses are the limiting values of the product $\rho \cdot (\dimensionless_time_coord - \elle)$ evaluated at $\elle$ numerically.}]{\includegraphics[width = 0.45\textwidth]{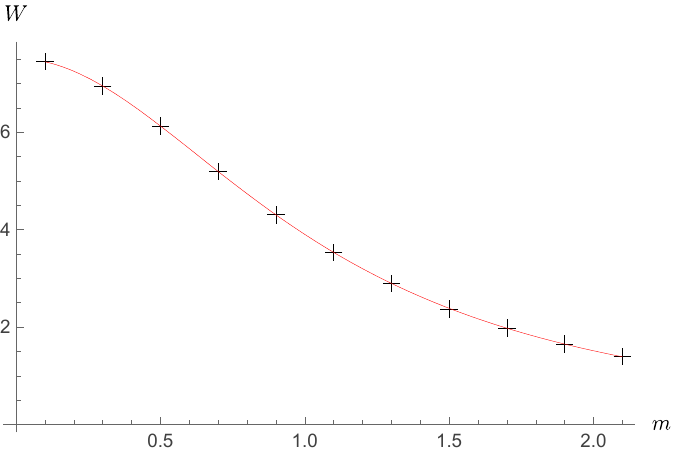}
}
\hspace{1em}
\subfloat[{To check the accuracy, we plot the difference between the theoretical value of $W$ and its numerical value, divided by $W$.}]{\includegraphics[width = 0.45\textwidth]{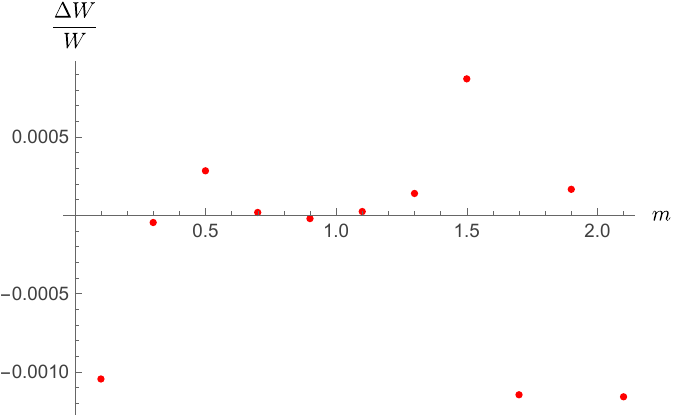}
}
\caption{The steepness of the singularity at $\elle$ given by the constant $W$ of (\ref{log_log_density_dependence}) as a function of $m$. The numerical values follow the theoretical expression.}
\label{the_W_constant}
\end{figure}

The density of the dust (\ref{density}) diverges to $+\infty$ at $\eta_0$ where the integral vanishes. Although the integral appearing in the density does not evaluate to a simple expression in general, we can still explore analytically the dust density shortly after the singularity that occurs at the time $\eta = \eta_0$. The integral is an increasing function here since its integrand is positive, so that we can Taylor expand it as 
\begin{equation}\label{Taylor_expansion_of_the_integral}
  \int_{\eta_0}^\eta  \frac{\xi^2}{f(\xi)^\frac{3}{2}} \; \mathrm{d}\xi \sim \frac{\eta_0^2}{f(\eta_0)^{\frac{3}{2}}} (\eta - \eta_0),
\end{equation}
to obtain an expression for the diverging dust density
\begin{equation}\label{dust_density_near_the_singularity}
\rho \sim \frac{1}{4 \pi} \frac{f(\eta_0)}{\eta_0^3 (\eta - \eta_0)}.
\end{equation}
Therefore, we get the relation
\begin{equation}\label{log_log_density_dependence}
    \log \rho \sim - \log (\eta -\eta_0) + \log \frac{\frac{\Lambda}{3} \eta_0^4 + \eta_0^2 + \alpha \eta_0 - M^2}{4 \pi \eta_0^3} =: - \log (\eta -\eta_0) + \log W,
\end{equation}
where we defined a new constant $W = (\frac{\Lambda}{3} \eta_0^4 + \eta_0^2 + \alpha \eta_0 - M^2)/(4 \pi \eta_0^3)$ that determines the steepness of the singularity. This relation is confirmed by \autoref{log_log_plot_density_only_alpha_and_M} and \autoref{the_W_constant}. The same dependence also applies to the Ricci scalar by virtue of the relation (\ref{Ricci_scalar}). Near the singularity, the proper time of comoving observers goes as
\begin{equation}
    \proper_time_coord \sim \frac{\eta_0}{\sqrt{f(\eta_0)}} \left( \eta - \eta_0 \right),
\end{equation}
so that
\begin{equation}\label{dust_density_near_the_singularity_in_proper_time}
\rho \sim \frac{1}{4 \pi} \frac{\sqrt{f(\eta_0)}}{\eta_0^2 \proper_time_coord} = \frac{1}{4 \pi} \frac{\sqrt{{\frac{\Lambda}{3} \eta_0^4 + \eta_0^2 + \alpha \eta_0 - M^2}}}{\eta_0^2 \proper_time_coord}.
\end{equation}
The energy density of the magnetic field (\ref{density}) approaches a finite value
\begin{equation}\label{magnetic_field_energy_density_at_the_singularity}
    \rho_{\mathrm{B}} = \frac{M^2}{8 \pi \eta_0^4}.
\end{equation}
If $\Lambda=0$ then we already know the value of $M$ from the asymptotic behavior as $t \rightarrow \infty$ given by (\ref{magnetic_field_energy_asymptotics}). Assuming $M \not = 0$, we are then able to determine the value of $\eta_0$ from (\ref{magnetic_field_energy_density_at_the_singularity}). Once we have $\eta_0$, we return to (\ref{dust_density_near_the_singularity_in_proper_time}) to find the value of the last missing parameter $\alpha$, which determines the steepness of the singularity. If the magnetic field is identically zero at all times with $M=0$ or $\Lambda \not = 0$ then (\ref{magnetic_field_energy_density_at_the_singularity}) does not allow us to derive the value of $\eta_0$ and we need to go two orders higher in the Taylor expansion near the singularity (the first two orders contain the same combination of the metric parameters). This yields
\begin{equation}\label{dust_density_near_the_singularity_in_proper_time_to_the_second_order}
\rho \sim \frac{1}{4 \pi} \frac{\sqrt{f(\eta_0)}}{\eta_0^2 \proper_time_coord} - \frac{3}{8 \pi} \frac{f(\eta_0)}{\eta_0^4} + \frac{\sqrt{f(\eta_0)} \left( -29 M^2 + 27 \alpha \eta_0 + 23 \eta_0^2 + 
  3 \eta_0^4 \Lambda \right)}{48 \pi \eta_0^6} t,
\end{equation}
consistent with the particular case of Section \ref{Lambda = 0, M = 0, alpha not= 0}. We now have (\ref{magnetic_field_energy_density_at_the_singularity}) and (\ref{dust_density_near_the_singularity_in_proper_time_to_the_second_order}) that enable us to determine the remaining 3 parameters of the metric $\alpha, \eta_0,$ and $M$. Therefore, these parameters fix the primordial value of the magnetic field and the steepness and form of the initial singularity of the dust density.
\subsection{Factorization of the master function}\label{factorization}
\begin{figure}[h]
\centering
\subfloat[The rescaled master function $\Phi(\dimensionless_time_coord)$ (\ref{rescaled_master_function_in_terms_of_rescaled_constants}). The larger root is located approximately at $\dimensionless_time_coord \approx 0.57$ where the integral diverges.]{\includegraphics[width = 0.45\textwidth]{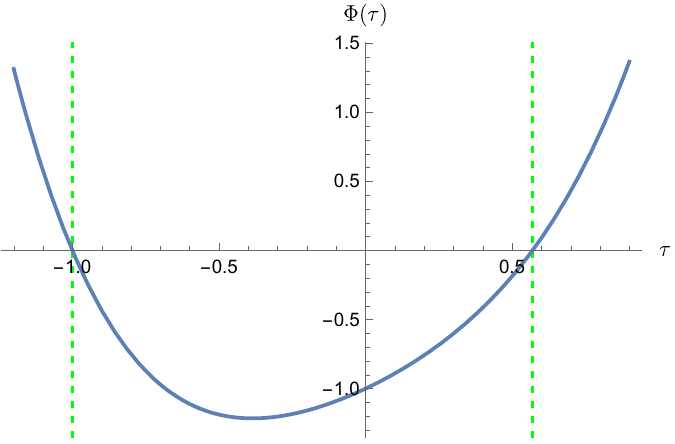}\label{plot_master}}
\hspace{1em}
\subfloat[The integral $\Psi(\dimensionless_time_coord) = \int_\elle^\dimensionless_time_coord \left( \zeta^2 /\rescaled_master_function(\zeta)^{\frac{3}{2}} \right) \mathrm{d}\zeta$ with a particular choice of the additive integration constant fixing the asymptotic value of the integral to be $\beta \approx 0.18 > 0$ and determining $\dimensionless_time_coord_0 \approx 0.95$ where the integral vanishes and where there is a singularity of the spacetime.]{\includegraphics[width = 0.45\textwidth]{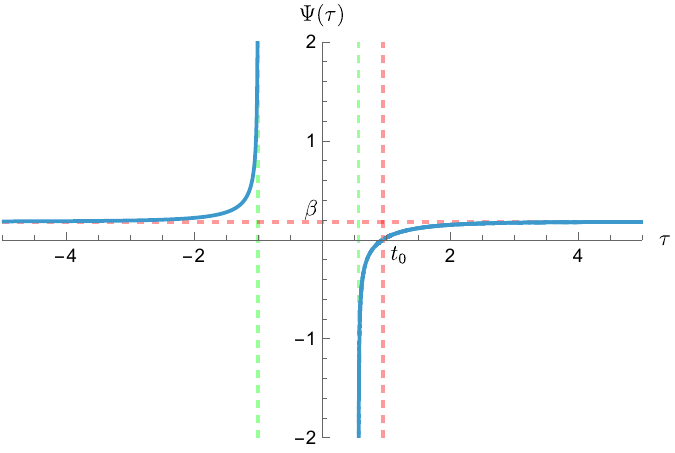}\label{plot_integral}}
\caption{In these plots we set $\lambda=1, m=1$. The two real roots of the master function $\rescaled_master_function(\dimensionless_time_coord)$ are denoted by the dashed green vertical lines. The integral appearing in (\ref{rescaled_metric}) is only real outside of the time interval between the roots. The dashed vertical red line denotes the point $\dimensionless_time_coord_0$ where the integral vanishes for a particular choice of the additive integration constant, which determines the asymptotic value of the integral $\beta$, denoted by the dashed horizontal red line.}
\end{figure}

\begin{figure}[h]
\centering
\subfloat[The density of the dust with $8 \pi \alpha^2 = 1$.]{\includegraphics[width = 0.45\textwidth]{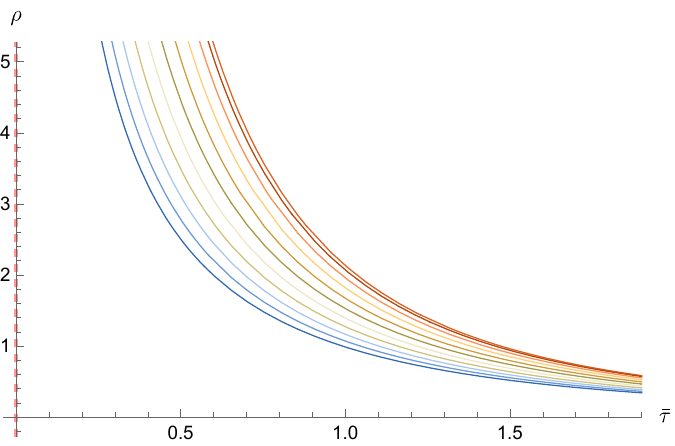}\label{dust_density_non_zero_beta_non_zero_lambda}}
\hspace{1em}
\subfloat[The ratio of the energy density of the magnetic field to that of the dust $\rho_{\mathrm{B}} / \rho$ (\ref{rescaled_density}).]{\includegraphics[width = 0.45\textwidth]{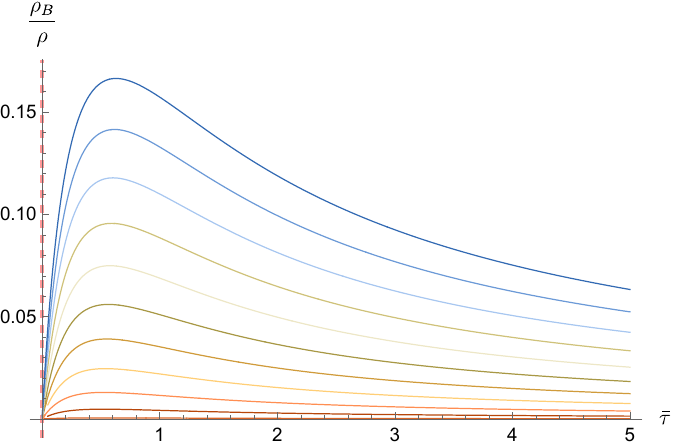}\label{magnetic_to_dust_energy_density_ratio_non_zero_beta_non_zero_lambda}}
\caption{In these plots corresponding to the general case, we use $\lambda = 1$, the asymptotic value $\beta \approx 0.18,$ and a shifted time coordinate $\bar{\dimensionless_time_coord} := \dimensionless_time_coord - \dimensionless_time_coord_0$ so that the singularity always occurs at $\bar{\dimensionless_time_coord}=0$. The curves correspond to $m = \{0.1, 0.3, 0.5, \ldots, 2.1\}$ from orange to blue. For higher $\beta$'s the ratio $\rho_{\mathrm{B}} / \rho$ can exceed 1 like in \autoref{magnetic_to_dust_energy_density_ratio_non_zero_beta}.}
\end{figure}

Let us now try to factorize the master function $\rescaled_master_function(\dimensionless_time_coord) = \lambda \dimensionless_time_coord^4 + \dimensionless_time_coord^2 + \dimensionless_time_coord - m^2$ to simplify the integral---that is, let us try to find a special set of the two parameters $(\lambda, m)$ for which we could perhaps evaluate the integral in (\ref{rescaled_metric}) explicitly. We want to explore the general case with both the cosmological constant and the magnetic field in the play: $\lambda, m \not = 0$. As argued above based on the signature of the metric, we need $\rescaled_master_function(\dimensionless_time_coord)>0$. We first notice that there are at least two distinct real solutions for $\lambda>0$, one positive and one negative. Also, none of the roots of the polynomial can be equal to zero due to the presence of $-m^2$. Now, what cannot be done:
\begin{itemize}
  \item $\rescaled_master_function(\dimensionless_time_coord) \not= \lambda (\dimensionless_time_coord-A)^4, \rescaled_master_function(\dimensionless_time_coord) \not= \lambda (\dimensionless_time_coord-A)(\dimensionless_time_coord-B)^3, \rescaled_master_function(\dimensionless_time_coord) \not= \lambda (\dimensionless_time_coord-A)^2(\dimensionless_time_coord-B)^2$ since all these can be ruled out algebraically by comparing the various powers of $\dimensionless_time_coord$ in the expansion.
  \item $\rescaled_master_function(\dimensionless_time_coord) \not= \lambda (\dimensionless_time_coord^2+A\dimensionless_time_coord+B)(\dimensionless_time_coord-C)^2$ with a single real double root since this requires $\lambda<0$ so that there is no interval of $\rescaled_master_function(\dimensionless_time_coord)>0$ as required by the signature of the metric.
  \item $\rescaled_master_function(\dimensionless_time_coord) \not= \lambda (\dimensionless_time_coord-A)(\dimensionless_time_coord-B)(\dimensionless_time_coord-C)^2,$ with two simple roots and one double root, since $C=-(A+B)/2, \lambda = -4/(3A^2 +2AB +3B^2)<0$ and $0<m^2 = AB(A+B)^2/(3A^2 +2AB +3B^2)$ (so that $AB>0$). We thus must have $C<0<A<B$ or $A<B<0<C$ (we can always rename $A$ and $B$ to reorder these two roots). The positive section of the master function is therefore delimited by the two simple roots $A$ and $B$. However, evaluating the proper time (\ref{proper_time}) above the smaller of the two simple roots $\rescaled_master_function(\dimensionless_time_coord) \sim \dimensionless_time_coord-A$, we get
      $$\proper_time_coord \sim \int \dimensionless_time_coord \frac{\mathrm{d}\dimensionless_time_coord}{\sqrt{\dimensionless_time_coord-A}} \sim \sqrt{\dimensionless_time_coord-A} \;,$$
      which is finite (the same applies below the larger simple root with $\sqrt{B-\dimensionless_time_coord}$). Therefore, the total interval of the proper time of a comoving observer between the two roots is finite. The solution thus must be extended beyond the physically acceptable interval but that would produce the wrong signature due to the negative sign of $\rescaled_master_function(\dimensionless_time_coord)$.
  \item $\rescaled_master_function(\dimensionless_time_coord) \not= \lambda (\dimensionless_time_coord-A)(\dimensionless_time_coord-B)(\dimensionless_time_coord-C)(\dimensionless_time_coord-D),$ with four simple real roots, since if we require $\lambda>0$ as per the arguments above, we cannot factorize the master function $f$ in this manner.

\end{itemize}
So the only feasible case is
\begin{equation}\label{factorization_of_f}
  \rescaled_master_function(\dimensionless_time_coord) = \lambda (\dimensionless_time_coord-A)(\dimensionless_time_coord-B)(\dimensionless_time_coord^2+C\dimensionless_time_coord+D)
\end{equation}
with two simple real roots $A<0<B$. We could, in principle, have one of the simple roots located at $\dimensionless_time_coord=0$, changing the properties of the integral in (\ref{the_metric}), but that would require $m^2 = 0$ which we exclude since we do want to have the magnetic field in the solution. Therefore, this is the only physically plausible factorization. The integral appearing in (\ref{rescaled_metric}) can then  be evaluated exactly: it is an expression involving elliptic integrals due to the last bracket in (\ref{factorization_of_f}), which is too complicated to give in the main text---we present it in Appendix (\ref{Appendix}) instead. Here, we show the resulting plots of the master function (\autoref{plot_master}) and the integral itself (\autoref{plot_integral}) for a particular choice $\lambda=1, m=1$, entailing $A=-1, B \approx 0.57, C \approx - 0.43, D \approx 1.75$. We also present the density of the dust (\autoref{dust_density_non_zero_beta_non_zero_lambda}) and the ratio of energy density of the dust and of the magnetic field (\autoref{magnetic_to_dust_energy_density_ratio_non_zero_beta_non_zero_lambda}) for $\lambda = 1$, the asymptotic value $\beta \approx 0.18$, and a range of $m$'s. To the best of our knowledge, an explicit solution of the general case has never been presented in the literature so far. On the other hand, it is quite a long expression and it is thus perhaps more useful to explore it by plotting the resulting physical characteristics of the spacetime as we chose here.

By choosing a negative asymptotic value $\beta<0$ of the integral appearing in (\ref{rescaled_metric}), we can make it vanish at some negative $\dimensionless_time_coord_0<0$. Then the density is positive below this value and the proper time diverges logarithmically as $\dimensionless_time_coord \rightarrow -\infty$. The situation is thus analogous to the one described in the above plots only that instead of starting at a singularity and extending the expanding solution to infinite times, we start in the asymptotic area and proceed with a collapse to the singularity.
\subsection{The influence of the magnetic field}\label{comparison}
\begin{figure}[h]
\centering
\includegraphics[width = 0.7\textwidth]{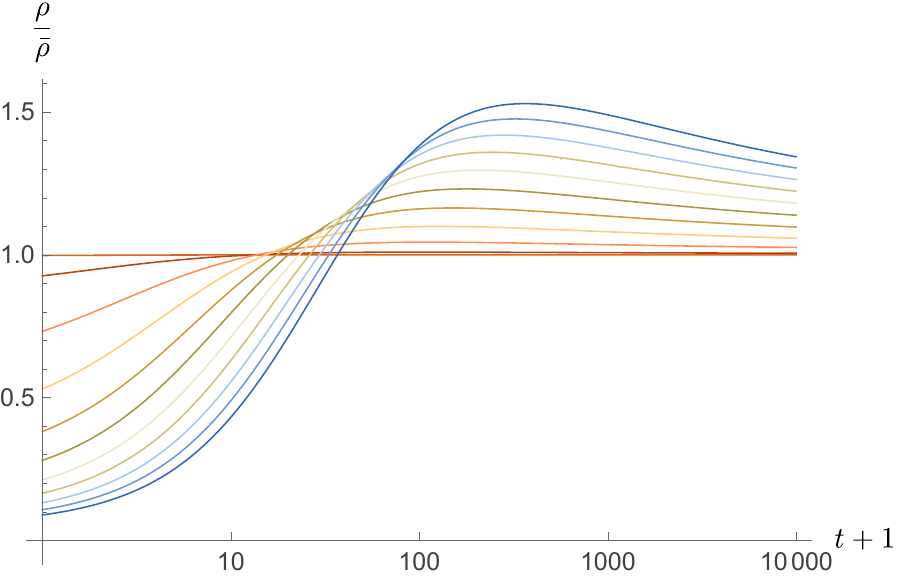}
\caption{The ratio of the dust density with the magnetic field present, $\rho$ of (\ref{Density_only_alpha_and_M}), to the dust density without the magnetic field, $\bar{\rho}$ of (\ref{density_only_alpha}), as a function of the proper time, $t$, elapsed from the moment of singularity with the integration constant $\beta = -1$ in both cases and $m = \{ 0.1, 0.3, 0.5, \ldots, 2.1 \}$, from orange to blue. Notice the logarithmic scaling of the horizontal axis and the fact that it is shifted by 1 so that the singularity occurs at the vertical axis.}\label{comparison_plot}
\end{figure}

To judge the influence of the presence of the magnetic field on the properties of the spacetime, we evaluated the ratio of the dust densities with and without the magnetic field, defined by the relations (\ref{Density_only_alpha_and_M}) and (\ref{density_only_alpha}), respectively. We plotted the ratio as a function of the proper time of comoving observers. However, we must bear in mind that we are comparing two different spacetimes, not two scalars within a single spacetime. To focus on the magnetic field only, we set the cosmological constant equal to zero. The resulting plot in \autoref{comparison_plot} shows that, as expected, the two densities are almost equal for weak magnetic fields (the horizontal orange line). At the singularity, the ratio starts from a finite value below 1, increases above 1, and ultimately settles back to 1 at late times. The finite value of the ratio at the singularity is due to the relation (\ref{dust_density_near_the_singularity}), which is valid for both densities. The fact that the ratio approaches 1 for late times results from the two spacetimes having the same asymptotics as discussed in Sections (\ref{Lambda = 0, M = 0, alpha not= 0}) and (\ref{Lambda = 0, M not= 0, alpha not= 0}). From this perspective, the magnetic field lowers the dust density in the beginning after the singularity and then, from some moment on, increases it as compared to a spacetime without the magnetic field. Nevertheless, we need to remember that the relative importance of the magnetic field compared to that of the dust \emph{within} any given spacetime is limited, as illustrated by \autoref{magnetic_to_dust_energy_density_ratio_non_zero_beta}.
\section{Conclusions}\label{Conclusions}
In this paper we presented and discussed several solutions of Einstein-Maxwell equations due to a gravitationally coupled magnetic field and pressureless dust and also the cosmological constant. The density of the dust in these solutions is positive so that the energy conditions are satisfied. The spacetime starts with a curvature singularity with the dust density diverging and then it expands indefinitely with the proper time of comoving observers becoming infinite and the solution approaching the de Sitter spacetime. Since these solutions are spatially homogeneous, all physical characteristics are functions of time only.

More precisely, we gave a new exact non-vacuum solution (\ref{I}) of Bianchi type III with pressureless dust and a magnetic field aligned with the axis of symmetry. The solution involves 4 parameters, which we denote $\alpha, \eta_0, \Lambda,$ and $M$: $\Lambda$ is the cosmological constant dictating the evolution of the model for large times as it approaches the de Sitter spacetime, or the line element (\ref{the_metric_everything_zero}) if $\Lambda=0$. The constant $M$ appears as an integration constant in Maxwell equations and determines the strength of the magnetic
field observed by comoving observers. Its value fixes the asymptotic fading of the energy density of the magnetic field for large proper times (\ref{magnetic_field_energy_asymptotics_Lambda}), or (\ref{magnetic_field_energy_asymptotics}) depending on the value of $\Lambda$. The parameter $\eta_0$ sets the moment of the initial singularity with the dust density (\ref{density}) diverging while the energy density of the magnetic field (\ref{magnetic_field_energy_density_at_the_singularity}) remains finite, corresponding thus to a primordial field. The remaining parameter of the spacetime, $\alpha$, together with the other 3 parameters determines the behavior of the dust density shortly after the initial singularity, which follows the relation (\ref{dust_density_near_the_singularity_in_proper_time_to_the_second_order})---it thus fixes the steepness and form of the initial singularity.

Therefore, there is no single quantity that we could measure at a particular time and that would in itself yield the value of a particular metric parameter---and vice versa: a single metric parameter does not fix any specific measurable quantity. In the general case only the cosmological constant can be determined in a straightforward manner by looking at the asymptotic value of the Ricci scalar, while to find the values of the other 3 parameters we need to follow the evolution of the dust and magnetic field densities in time. However, the explicit form of any quantity is rather unwieldy as can be seen in Appendix \ref{Appendix}. As suggested above, it seems easiest to study the rate of divergence of the dust density at the initial singularity and the primordial value of the magnetic field energy density. These yield three algebraic relations for the three sought parameters. The previous literature mostly just gave a list of the mathematical forms of possible solutions without discussing their physics.

We come to the conclusion that the magnetic field plays only a minor role in the evolution of the spacetime near the initial singularity at $\dimensionless_time_coord=\dimensionless_time_coord_0$ where its finite energy density proportional to the Maxwell invariant (\ref{rescaled_electromagnetic_field}) is swamped by the diverging density of the dust (\ref{rescaled_density}). Likewise, the magnetic field is less relevant asymptotically, in the limit $\dimensionless_time_coord \rightarrow \infty$: the Maxwell invariant goes as $F^{ \alpha \beta} F_{ \alpha \beta} \sim \dimensionless_time_coord^{-4}$ while the dust density goes as $\rho \sim \dimensionless_time_coord^{-3}$ (we must bear in mind that asymptotically, the integral in (\ref{rescaled_metric}) goes to a constant) so that the dust dissolves more slowly than the magnetic field. However, as we established in Section (\ref{asymptotics}), it is the cosmological constant that dominates here. We can thus distinguish three eras of this universe: the dust era after the initial singularity when the evolution is dictated by the density of the dust; this is then followed (for some choices of the parameters, see \autoref{magnetic_to_dust_energy_density_ratio_non_zero_beta}) by a magnetic-field era dominated by the energy density of the magnetic field; and ultimately, the universe enters a de Sitter era when the evolution is dominated by the cosmological constant. We also assessed the impact of the magnetic field on the properties of the spacetime, comparing the dust density in two different spacetimes differing only in the strength of the magnetic field. It turns out that the ratio of the two dust densities tends to a constant value towards the initial singularity. This constant is different from 1 and thus, in these models, the magnetic field does influence also the properties of the early universe.

In this paper, the magnetic field is a solution of the sourceless Maxwell equations so it is a self-consistent solution requiring no charges just like in the case of the Bonnor-Melvin spacetime \cite{Bonnor,Melvin}. These results can be generalized by, for example, considering a non-vanishing pressure of the fluid or assuming a negative cosmological constant, $\Lambda<0$. Although current observations do not favor large-scale charge distributions present in the universe, one can give the dust a charge density and make it the source of the electromagnetic field. If we combine two streams of dust with opposite charge densities we still get a vanishing total charge density. Thus, in our view, there is room for further interesting toy models exploring these options.
\section{Acknowledgments}
M.Z. is grateful for the support of grant GACR 22-14791S and wishes to thank Tomáš Ledvinka for fruitful discussions. Plots in the paper were generated by the Wolfram Mathematica software. 
\appendix
\section{The full integral}\label{Appendix}
Based on the discussion of possible factorizations of the master function $\rescaled_master_function(\dimensionless_time_coord)$, we need to evaluate the expression
\begin{equation}\label{general_integral}
\int_\elle^\dimensionless_time_coord \frac{\zeta^2 \; \mathrm{d}\zeta }{\rescaled_master_function(\zeta)^{\frac{3}{2}}} = \int_\elle^\dimensionless_time_coord \frac{\zeta^2 \; \mathrm{d}\zeta }{\left[ (\zeta+A)(\zeta-B)(\zeta^2+C \zeta+D) \right]^{\frac{3}{2}}},
\end{equation}
with $A,B>0$ and $C^2-4D<0$. Defining the root of the quadratic factor in the master function $2r:=-C+\sqrt{C^2-4D} =-C+i \sqrt{|C^2-4D|}$, we find an explicit form of the integral (\ref{general_integral}):
\begin{eqnarray}\label{I}
  \int_\elle^\dimensionless_time_coord \frac{\zeta^2 \; \mathrm{d}\zeta }{\rescaled_master_function(\zeta)^{\frac{3}{2}}} &=& \beta + \frac{2 (\mathfrak{a} + \mathfrak{b}\dimensionless_time_coord+ \mathfrak{c} \dimensionless_time_coord^2)}{(A+B) (A+r) (A+\bar{r}) (B-r)^2 (B-\bar{r})^2 (r-\bar{r})^2 \sqrt{\rescaled_master_function(\dimensionless_time_coord)}} \nonumber\\
  && + \frac{\mathfrak{A} }{\sqrt{(r+A)(B-\bar{r})}} \; F\left(\sin ^{-1} \sqrt{\frac{(\dimensionless_time_coord+A)(r-B)}{(\dimensionless_time_coord-B)(r+A)}} ; \sqrt{\frac{(r+A)(\bar{r}-B)}{(r-B)(\bar{r}+A)}}\right) \nonumber\\
  && + \mathfrak{B} \; \sqrt{(r+A)(B-\bar{r})} \; E\left(\sin ^{-1} \sqrt{\frac{(\dimensionless_time_coord+A)(r-B)}{(\dimensionless_time_coord-B)(r+A)}} ; \sqrt{\frac{(r+A)(\bar{r}-B)}{(r-B)(\bar{r}+A)}}\right),
\end{eqnarray}
with $\beta$ an additive integration constant and
\begin{eqnarray}
  \mathfrak{a} &=& -B r \bar{r} \left(\right. A^2 \left(B^2 (r+\bar{r})+B \left(r^2-6 r \bar{r}+\bar{r}^2\right)+r \bar{r} (r+\bar{r})\right) \nonumber\\
  &&+A \left(B^3 (r+\bar{r})-2 B^2 r \bar{r}+B \left(r^3-2 r^2 \bar{r}-2 r \bar{r}^2+\bar{r}^3\right)+r \bar{r} \left(r^2+\bar{r}^2\right)\right) \nonumber\\
  && +2 B r \bar{r} \left(B^2-B (r+\bar{r})+r^2-r \bar{r}+\bar{r}^2\right) \left.\right), \\
  \mathfrak{b} &=& A^2 \left(B^3 \left(r^2+\bar{r}^2\right)+B^2 \left(r^3-2 r^2 \bar{r}-2 r \bar{r}^2+\bar{r}^3\right)-2 B r^2 \bar{r}^2+r^2 \bar{r}^2 (r+\bar{r})\right) \nonumber\\
  && +A \left(B^4 \left(r^2+\bar{r}^2\right)+B^2 \left(r^4-6 r^2 \bar{r}^2+\bar{r}^4\right)+r^2 \bar{r}^2 \left(r^2+\bar{r}^2\right)\right) \nonumber\\
  && +B r \bar{r} \left(B^3 (r+\bar{r})-2 B^2 r \bar{r}+B \left(r^3-2 r^2 \bar{r}-2 r \bar{r}^2+\bar{r}^3\right)+r \bar{r} \left(r^2+\bar{r}^2\right)\right), \\
  \mathfrak{c} &=& -2 A^2 \left(B^2 \left(r^2-r \bar{r}+\bar{r}^2\right)-B r \bar{r} (r+\bar{r})+r^2 \bar{r}^2\right) \nonumber\\
  && -A \left(B^3 \left(r^2+\bar{r}^2\right)+B^2 \left(r^3-2 r^2 \bar{r}-2 r \bar{r}^2+\bar{r}^3\right)-2 B r^2 \bar{r}^2+r^2 \bar{r}^2 (r+\bar{r})\right) \nonumber\\
  && -B r \bar{r} \left(B^2 (r+\bar{r})+B \left(r^2-6 r \bar{r}+\bar{r}^2\right)+r \bar{r} (r+\bar{r})\right), \\
  \mathfrak{A} &=& \frac{2}{(A+B)^2 (A+r) (A+\bar{r}) (B-r) (B-\bar{r}) (r-\bar{r})^2} \Bigg[ A^3 \left(r \bar{r} (r+\bar{r})-B \left(r^2+\bar{r}^2\right)\right) \nonumber\\
  && +A^2 \left(-4 B^2 r \bar{r}-B \left(r^3-2 r^2 \bar{r}-2 r \bar{r}^2+\bar{r}^3\right)+r \bar{r} \left(r^2+\bar{r}^2\right)\right) \nonumber\\
  && +A B \left(-\left(B^2 \left(r^2+\bar{r}^2\right)\right)+B \left(r^3-2 r^2 \bar{r}-2 r \bar{r}^2+\bar{r}^3\right)+4 r^2 \bar{r}^2\right) \nonumber\\
  && +B^2 r \bar{r} \left(-B (r+\bar{r})+r^2+\bar{r}^2\right) \Bigg], \\
  \mathfrak{B} &=& \frac{4}{(A+B)^2 (A+r)^2 (A+\bar{r})^2 (B-r)^2 (B-\bar{r})^2 (r-\bar{r})^2} \Bigg[ \nonumber\\
  && A^4 \left(B^2 \left(r^2-r \bar{r}+\bar{r}^2\right)
    -B r \bar{r} (r+\bar{r})+r^2 \bar{r}^2\right) \nonumber\\
  && +A^3 \left(B^3 \left(r^2+\bar{r}^2\right)+B^2 \left(r^3-2 r^2 \bar{r}-2 r \bar{r}^2+\bar{r}^3\right)-2 B r^2 \bar{r}^2+r^2 \bar{r}^2 (r+\bar{r})\right) \nonumber\\
  && +A^2 \left(\right. B^4 \left(r^2-r \bar{r}+\bar{r}^2\right)-B^3 \left(r^3-2 r^2 \bar{r}-2 r \bar{r}^2+\bar{r}^3\right) \nonumber\\
  && +B^2 \left(r^4+2 r^3 \bar{r}-12 r^2 \bar{r}^2+2 r \bar{r}^3+\bar{r}^4\right)-B r \bar{r} \left(r^3-2 r^2 \bar{r}-2 r \bar{r}^2+\bar{r}^3\right) \nonumber\\
  && +r^2 \bar{r}^2 \left(r^2-r \bar{r}+\bar{r}^2\right) \left.\right)+A B r \bar{r} \left(\right. B^3 (r+\bar{r})-2 B^2 r \bar{r}+B \left(r^3-2 r^2 \bar{r}-2 r \bar{r}^2+\bar{r}^3\right) \nonumber\\
  && +r \bar{r} \left(r^2+\bar{r}^2\right) \left.\right)+B^2 r^2 \bar{r}^2 \left(B^2-B (r+\bar{r})+r^2-r \bar{r}+\bar{r}^2\right) \Bigg].
\end{eqnarray}
The above expressions are in general complex. However, the imaginary part is constant below and above the lower and upper real roots, $-A$ and $B$, respectively, while the real part is constant between the two real roots. Choosing an appropriate integration constant $\beta$, we can thus make the resulting function real below and above the roots, and purely imaginary between them. This is in line with the discussion below relation (\ref{the_metric}). The choice of the sign of the real part of $\beta$ and its consequences are discussed in Section \ref{properties_of_the_integral} and the general form of the resulting integral is shown in Figure (\ref{plot_integral}).
\newpage
\bibliographystyle{IEEEtran}
\bibliography{Bianchi_III}{}
\end{document}